\documentclass[notitlepage,lengthcheckaps,onecolumn,showpacs,floatfix,superscriptaddress]{revtex4-1}

\usepackage{amsmath,amsfonts,amssymb,amsthm}
\usepackage{graphicx,svg}
\usepackage{hyperref}
\usepackage{float}
\usepackage{color}
\usepackage{braket}
\usepackage{mathrsfs}
\usepackage{thmtools}
\usepackage{mathtools}
\usepackage{todonotes}
\usepackage{caption}

\DeclarePairedDelimiter{\floor}{\lfloor}{\rfloor}

\theoremstyle{plain}

\theoremstyle{definition}

\theoremstyle{remark}

\def\be{\begin{equation}}
\def\ee{\end{equation}}
\def\ba{\begin{eqnarray}}
\def\ea{\end{eqnarray}}
\def\lo{\longrightarrow}
\def\h{\hskip 1cm }

\def\la{\langle}
\def\ra{\rangle}
\def\a{\alpha}

\begin{document}

\title{Planar Maximally Entangled States}
\author{Mehregan Doroudiani }\email{mehregandoroudiani@gmail.com}

\affiliation{Department of Physics, Sharif University of Technology, P.O.Box 11155-9161, Tehran, Iran.}
\author{Vahid Karimipour}\email{vahid@sharif.edu}
\affiliation{Department of Physics, Sharif University of Technology, P.O.Box 11155-9161, Tehran, Iran.}

\begin{abstract}
We construct a large family of  Planar Maximally Entangled (PME) states which are a wider class of multi-partite entangled states than Absolutely Maximally Entangled (AME) states.  These are states in which any half of the qudits are in a maximally mixed state, provided that they form a connected subset.  We show that in contrast to AMEs, PMEs are easier to find and there are various PMEs for any even number of qudits. In particular, while it is known that no AME state of four qubits exists,  we show that there are  two distinct multi-parameter classes of four qubit PMEs.  We also give explicit families of PMEs for any even number of particles and for any dimension. We also briefly mention the relevance of connectivity and  underlying geometry to definition of a PME state. Here we have considered the simplest case of a one dimensional graph.  \end{abstract}

\maketitle

\section{Introduction} \label{int}
In contrast to bipartite entangled states where we have a complete knowledge about their characterization and measures of entanglement  at least for pure states, our knowledge about multipartite states is far from satisfactory. It is now well known that multipartite entanglement is an important resource in areas like quantum networks \cite{peres}, tensor networks \cite{tensor},  quatum metrology \cite{maccone1, metro, noise}, and distributed quantum computing \cite{rausendorf, blind}. This makes the study of multipartite entanglement  an appealing and important problem in quantum information theory which may remain to be so for years. While for the bipartite case, Bell states define the top maximally entangled states in the hierarchy of bipartite states, for the multi-partite states there is no single hierarchy and it has been shown that there are independent families of states which cannot be inter-converted by Local Operation and Classical Communication (LOCC) \cite{ciracGHZ, entReview1, entReview2}. The number of these families grows rapidly with the number of particles. Moreover, in the multipartite case the number of possible partitions for which we have to characterize the entanglement grows exponentially, making the problem of defining maximally entangled states very problematic. Therefore one has to define maximal entanglement with a view toward specific applications. It is in this background that in recent years the concept of Absolutely Maximally Entangled (AME) states has emerged and has been extensively studied \cite{helwig2012absolute, helwig2013absolutely, helwig2,  ame5, raissi1, raissi2}. These are the states of $n$ particles such that any collection of $\floor*{\frac{n}{2}} $ particles are in a maximally mixed state. A prototype of this is the GHZ state for 3 particles. It has been shown that these states have a rich structure with many  connections to other concepts in mathematics and applications to physics  including  parallel teleportation and   threshold quantum secret sharing schemes \cite{helwig2012absolute}, 
quantum error correcting codes\cite{raissi1}, combinatorial designs \cite{carol1,carol2,carol3}, and microscopic models \cite{Almheiri_2015,harlow2018tasi,pastawski2015holographic} of holographic duality \cite{Maldacena:1997re,Witten_1998}. \\

\noindent In a long search for the AME states, it has been shown that for each fixed number of particles $n$ when the dimension $d$ of states grows, it is generally possible to find such states, however for low dimensions there are severe restrictions. For example it has been shown that for qubits ($d=2$), there are no AME states when $n=4$ \cite{higuchi}. For 
$n=5$ and $n=6$ AME states are known to exist \cite{amelist1, amelist2, amelist3}, and $n=7$ qubit states are proved not to exist \cite{ame7qubit}.  See \cite{List} for a list of positive and negative results. \\

\noindent Given the multitude of constraints that a state must satisfy in order to be an AME state and the subsequent rarity of such states, one may ask if it is possible to search for a wider class of entangled states with a lower number of constraint. At first this may be simply a mathematical curiosity but it may be the case that  such states find applications as well. It was in \cite{berger2018perfect} that with this view, the notion of perfect tangles was put forward. These new states which for the sake of uniformity of nomenclature, we call Planar Maximally Entangled (PME) states have the property that any collection of $\floor*{\frac{n}{2}}$ {\it {adjacent}} particles are in a completely mixed state and by adjacent here we also imply periodic boundary conditions. Figure (\ref{PMEPartition}) elaborates this concept. It is quite natural to expect there may be many more PME states than AME states. Nevertheless, we emphasize that finding such states is still a highly nontrivial task. In this work, we study these states and provide several classes of examples. In section (\ref{4p}), we show that in the four qubit case, there are two multi-parameter classes of  PMEs in contrast to the AME states which do not exist for four qubits.  Then in section (\ref{pmeN}), we introduce larger classes of such states for larger number of particles and for higher dimensional states. Finally in section (\ref{pmeApp}), we briefly discuss applications of these states in  different areas like teleportation  and quantum state sharing \cite{helwig2012absolute}.

\section{Some preliminary facts}\label{pre}
\noindent  Consider a partition of the particles into parts $A$ and $B$ with $|A|\leq |B|$ and expand a state $|\Psi\ra$ as
\be\label{basicPsi}
|\Psi\ra=\sum_{{\bf k}}|{\bf k}\ra_A|\phi_{\bf k}\ra_B.
\ee
Then the density matrix of part $A$ is given by
\be\label{basicRho}
\rho_A=\sum_{{\bf k},{\bf k'}}|{\bf k}\ra_A\la{\bf k'} |\la \phi_{\bf k}|\phi_{\bf k'}\ra_B.
\ee
For this density matrix to be maximally mixed, we require that the states $\{|{\bf k}\ra\}$ form a basis for the Hilbert space of part $A$ and the states $\{|\phi_{\bf k}\ra\}$ be orthogonal and with equal norm. Another way of saying this is to demand that the operator 

\be\label{basicOp}
\hat{\Psi}:=\sum_{{\bf k}}|\phi_{\bf k}\ra_B\la {\bf k}|_A: H_A\lo H_B,
\ee
be proportional to an  isometry, that is $\hat{\Psi}^\dagger \hat{\Psi}\propto I_{A}$. For AME states, this condition should be satisfied for all of the partitions, and for PME states, we demand that this condition be satisfied only for connected partitions, i.e. partitions composed of adjacent particles, given the periodic boundary condition.
When $|A|=|B|=N$, the operator $\hat{\Psi}$ will be proportional to a unitary. In this case, when 
$\rho_A=\frac{1}{N}I_A$, one has also $\rho_B=\frac{1}{N}I_B$. 
 Finally  we should note the obvious fact that when $A$ is in a maximally mixed state, any subset of $A$ is also in a maximally mixed states. 
 With these preliminaries, we can now  construct various families of PMEs for different even number of particles and in different dimensions. We start from the simplest case of four qubits.

\section{Four particle PME states for qubits }
\label{4p}
\noindent  For 3 particles, and in view of the periodic boundary conditions, there is no difference between an AME state and a PME state. The only AME state is the GHZ states $\ket{GHZ}=\frac{1}{\sqrt{2}}(\ket{000}+\ket{111})$. The difference between AME states and PME states starts at $n=4$ particles.  It is well known that for qubits, there is no AME states of four particles \cite{higuchi}. We now show that there are indeed PME states of qubits for $n=4$ and show that there are two distinct families of such states. \\

\noindent Consider a general state of four qubits

\be
|\Psi\ra=\frac{1}{2}\sum_{i,j,k,l=0}^{1}\a_{ijkl}|i,j,k,l\ra.
\ee
A basic result of \cite{sudbery} on generalized Schmidt decomposition of multi-partite states indicates that with a suitable choice of basis one can always fulfill:

\be \a_{1000}=\a_{0100}=\a_{0010}=\a_{0001}=0\ee.

\noindent  This state can be partitioned as 
\begin{equation}
\ket{\Psi}_{12,34} =\frac{1}{2}\Big( |00\ra|\phi_{00}\ra+|01\ra|\phi_{01}\ra+|10\ra|\phi_{10}\ra+|11\ra|\phi_{11}\ra\Big),
\end{equation}
where we have used the partition $(12,34)$ as denoted in the left hand side. For this state $\rho_{12}=\frac{I}{4}$, if and only if 
\be\label{norm}
\la \phi_{ij}|\phi_{kl}\ra=\delta_{ik}\delta_{jl}.
\ee
Expressed equivalently, this means that the matrix whose columns are the vectors $|\phi_{ij}\ra$ should be a unitary matrix. One can do a similar decomposition for the partition $(23,41)$ in the form
\begin{equation}
\ket{\Psi}_{23,41} =\frac{1}{2} \Big(|00\ra|\chi_{00}\ra+|01\ra|\chi_{01}\ra+|10\ra|\chi_{10}\ra+|11\ra|\chi_{11}\ra\Big),
\end{equation}
from which one concludes that the matrix whose columns are the vectors $\{|\chi_{ij}\ra\}$ should also be unitary. Expanding both sets of vectors in terms of the computational basis, one finds that the following different matrices formed out of the coefficients $\a_{ijkl}$ should be unitary at the same time:

\begin{equation}\label{UUU}
U = \left(\begin{array}{cccc}
{\alpha_{0000}} & {0} & {0} & {\alpha_{0011}} \\
{0} & {\alpha_{0101}} & {\alpha_{0110}} & {\alpha_{0111}} \\
{0} & {\alpha_{1001}} & {\alpha_{1010}} & {\alpha_{1011}} \\
{\alpha_{1100}} & {\alpha_{1101}} & {\alpha_{1110}} & {\alpha_{1111}}
\end{array}\right) ,
\end{equation}

\noindent  and

\begin{equation}\label{VVV}
W = \left(\begin{array}{cccc}
{\alpha_{0000}} & {0} & {0} & {\alpha_{0110}} \\
{0} & {\alpha_{1010}} & {\alpha_{1100}} & {\alpha_{1110}} \\
{0} & {\alpha_{0011}} & {\alpha_{0101}} & {\alpha_{0111}} \\
{\alpha_{1001}} & {\alpha_{1011}} & {\alpha_{1101}} & {\alpha_{1111}}
\end{array}\right) .
\end{equation}

\noindent {\bf Theorem:} There are two distinct classes of four-qubit PME states, these are a five parameter family 

\be\label{psiA}
|\Psi_{A}\ra_{1,2,3,4}=\frac{1}{2}\big[|0000\ra+e^{i\theta}|1010\ra+ e|0101\ra+d|1111\ra+g|0111\ra+m|1101\ra\big],
\ee
where 
$\left(\begin{array}{cc} e&g\\ m&d\end{array}\right)$ is a unitary matrix and another five-parameter family

\ba\label{psiB}
|\Psi_{B}\ra_{1,2,3,4}&=&\frac{1}{2}\cos\phi \big[e^{i\a}|0000\ra+e^{-i\a}|1111\ra+e^{i\beta}|0101\ra+e^{-i\beta}|1010\ra\big]\cr
&+& \frac{1}{2}\sin \phi \big[e^{i\gamma}|0011\ra-e^{-i\gamma}|1100\ra+e^{i\delta}|0110\ra-e^{-i\delta}|1001\ra\big].
\ea

\begin{figure}[H] 
	\centering
	\includegraphics[scale=0.5]{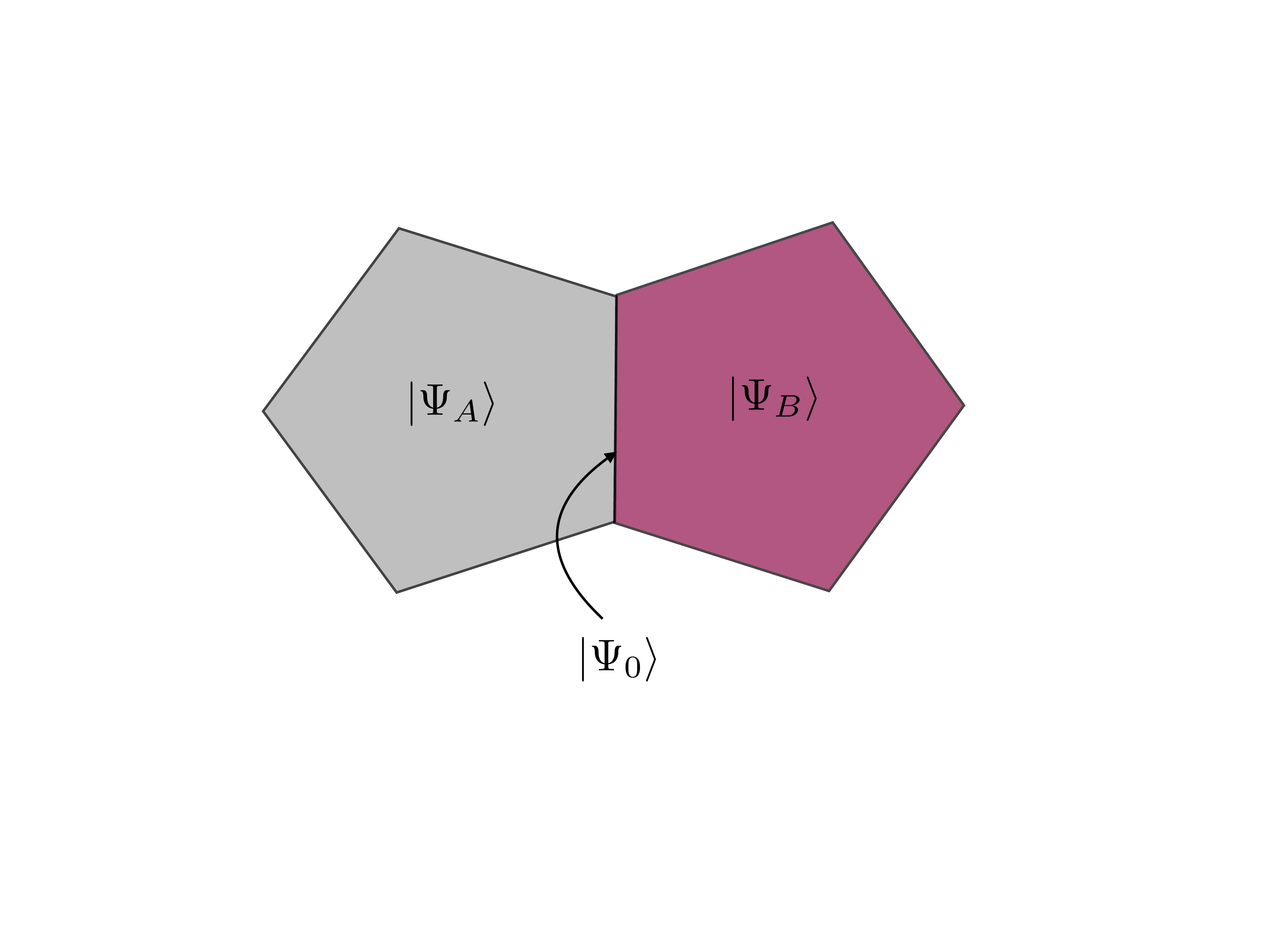}\vspace{-4.0cm}
	\caption{There are two 5-parameter families of 4-qubit PME states. They intersect each other at a one-parameter family given by $|\Psi_0\ra$. The number of edges symbolically depicts the number of parameter for each of the  three classes.}\label{pentagons}
\end{figure}

\noindent  The intersection of these two families is provided by setting $\phi=g=m=0$ and re-labeling the remaining phases, leading to the  one-parameter family below,  (see figure (\ref{pentagons}))

\be
|\Psi_0\ra=\frac{1}{2}\left[|0000\ra+|1111\ra+e^{i\beta}|0101\ra+
	e^{-i\beta}|1010\ra\right].
\ee

\noindent {\bf Proof:}
First we note that if $|\Psi\ra_{1,2,3,4}$ is a PME state, then its cyclic permutation $|\Psi\ra_{4,1,2,3}$ is also a PME state. We consider these two states equivalent. In order to classify all PME states of four qubits,  it is convenient to rename the entries of the two matrices $U$ and $V$ to simpler letters. So we want the following two matrices to be unitary:

\be\label{basic}
U= \left(\begin{array}{cccc}
	a & {0} & {0} & b \\
	{0} & e & f &g \\
	{0} & h & k & l \\
	c&m & n & d
\end{array}\right)\ \ \ \ \ {\rm and}\ \ \ \  W= \left(\begin{array}{cccc}
a & {0} & {0} & f \\
{0} & k & c &n\\
{0} & b & e & g \\
h&l& m & d
\end{array}\right),
\ee
 Orthogonality of rows and columns of both matrices, put many constraints on the remaining 12 complex  parameters. The problem is how to go through an exhaustive and exclusive list of this constraints which is also simple. To this end we start from the  condition $cm=0$ (from orthogonality of the first two columns of $U$) and then follow the consequences of our two choices ($c=0$ or $c\ne 0$). We repeatedly use the orthogonality of different rows and columns of both matrices,  use the property that if in any single row or column of a unitary matrix, an entry has unit modulus, then all the other entries should vanish. At the end we also use freedom of a global phase.  To make the argument transparent, we depict implications by right arrows $\rightarrow$ above which the letter  $U$ or $W$ indicates which matrix has been taken into account for that implication:\\

\noindent {\bf {\large Class A:}} $c=0$ \\

\noindent  We have from (\ref{basic})

\be
c=0\xrightarrow{U}a=1\xrightarrow{U} \ b=0, \ \ \  a=1\xrightarrow{W}(h=0\ \ f=0),
\ee
where we have used a global phase to put $a=1$. 
This leads to 

\be\label{basica}
U= \left(\begin{array}{cccc}
	1 & {0} & {0} & 0 \\
	{0} & e & 0&g \\
	{0} & 0 & k & l \\
	0&m & n & d
\end{array}\right)\ \ \ \ \ {\rm and}\ \ \ \  W= \left(\begin{array}{cccc}
	1 & {0} & {0} & 0 \\
	{0} & k & 0 &n\\
	{0} & 0 & e & g \\
	0&l& m & d
\end{array}\right),
\ee

\noindent We will now divide this class into two subclasses  as follows:\\

\noindent {\bf Subclass A1:}\ {$g\ne 0$.}\\

\noindent Then we have from (\ref{basica})
\be
g\ne 0\xrightarrow{U} l=0, \ {\rm and}\ \ \xrightarrow{W}n=0. 
\ee
which leads to the following form for the matrices

\be\label{basica1}
U= \left(\begin{array}{cccc}
	1 & {0} & {0} & 0 \\
	{0} & e & 0&g \\
	{0} & 0 & e^{i\theta} & 0 \\
	0&m & 0 & d
\end{array}\right)\ \ \ \ \ {\rm and}\ \ \ \  W= \left(\begin{array}{cccc}
	1 & {0} & {0} & 0 \\
	{0} & e^{i\theta} & 0 &0\\
	{0} & 0 & e & g \\
	0&0& m & d
\end{array}\right),
\ee
where $\left(\begin{array}{cc}
e & g  \\
m & d
\end{array}\right)$
is a unitary matrix with four parameters, this leads to the state (\ref{psiA}).\\

\noindent {\bf Subclass A2:}\ {$g= 0$.}\\

\noindent Then we  have from 
\be
g= 0\xrightarrow{U} e=e^{i\theta}, \ {\rm and}\ \ \xrightarrow{W}m=0. 
\ee
which leads to the following form for the matrices

\be\label{basica2}
U= \left(\begin{array}{cccc}
	1 & {0} & {0} & 0 \\
	{0} & e^{i\theta} & 0&0 \\
	{0} & 0 & k & l \\
	0&0 & n & d
\end{array}\right)\ \ \ \ \ {\rm and}\ \ \ \  W= \left(\begin{array}{cccc}
	1 & {0} & {0} & 0 \\
	{0} & k & 0 &n\\
	{0} & 0 & e^{i\theta} & 0 \\
	0&l& 0 & d
\end{array}\right),
\ee
where $\left(\begin{array}{cc}
k& n  \\
l & d
\end{array}\right)$
is a unitary matrix with four parameters. This leads to a state which is equivalent to the state (\ref{psiA}) with  a cyclic permutation. \\

\noindent{\bf Class B:} $c\ne 0$ \\

\noindent Then we have from (\ref{basic})

\be
c\ne 0 \xrightarrow{U}(m=0\ \ \ n=0),
\ee
\noindent This leads to the following form for the matrices

\be\label{basicb}
U= \left(\begin{array}{cccc}
	a & {0} & {0} & b \\
	{0} & e & f &g \\
	{0} & h & k & l \\
	c&0& 0 & d
\end{array}\right)\ \ \ \ \ {\rm and}\ \ \ \  W= \left(\begin{array}{cccc}
	a & {0} & {0} & f \\
	{0} & k & c &0\\
	{0} & b & e & g \\
	h&l& 0 & d
\end{array}\right).
\ee

\noindent {\bf Subclass B-1:}\ {$b\ne 0$.}\\

\noindent Then we  have from (\ref{basicb})
\be
b\ne 0\xrightarrow{U} (g=0, \ l=0),
\ee
which leads to the following form for the matrices

\be\label{basicb1}
U= \left(\begin{array}{cccc}
	a & {0} & {0} & b \\
	{0} & e & f&0 \\
	{0} & h & k & 0 \\
	c&0 & 0 & d
\end{array}\right)\ \ \ \ \ {\rm and}\ \ \ \  W= \left(\begin{array}{cccc}
	a & {0} & {0} & f \\
	{0} & k & c &0\\
	{0} & b & e & 0 \\
	h&0& 0 & d
\end{array}\right),
\ee
where the four inner and outer $2\times 2$ matrices should be unitary.
Parameterizing a general $2\times 2$ unitary matrix as 
$
\omega=\left(\begin{array}{cc}
	\cos\theta e^{i(\xi+\a)} & \sin\theta e^{i(\xi+\beta)
} \\
	 -\sin\theta e^{i(\xi-\beta)} & \cos\theta e^{i(\xi-\a)}
\end{array}\right)
$, imposing this form on all the four $2\times 2$ unitary matrices,
 solving the simple issuing equalities in different matrices and then finally using the freedom of a global phase, the final form of the resulting state turns out to be (\ref{psiB}).\\

\noindent{\bf Subclass B-2:}\ {$b= 0$.}\\

\noindent Then we  have from (\ref{basicb})
\be
b= 0\xrightarrow{U} (|a|=1)\xrightarrow{U} c=0, 
\ee
which contradicts the basic condition of this case, namely $c\ne 0$. No new state results. \\

\begin{figure}[H] 
	\centering
	\includegraphics[scale=0.2]{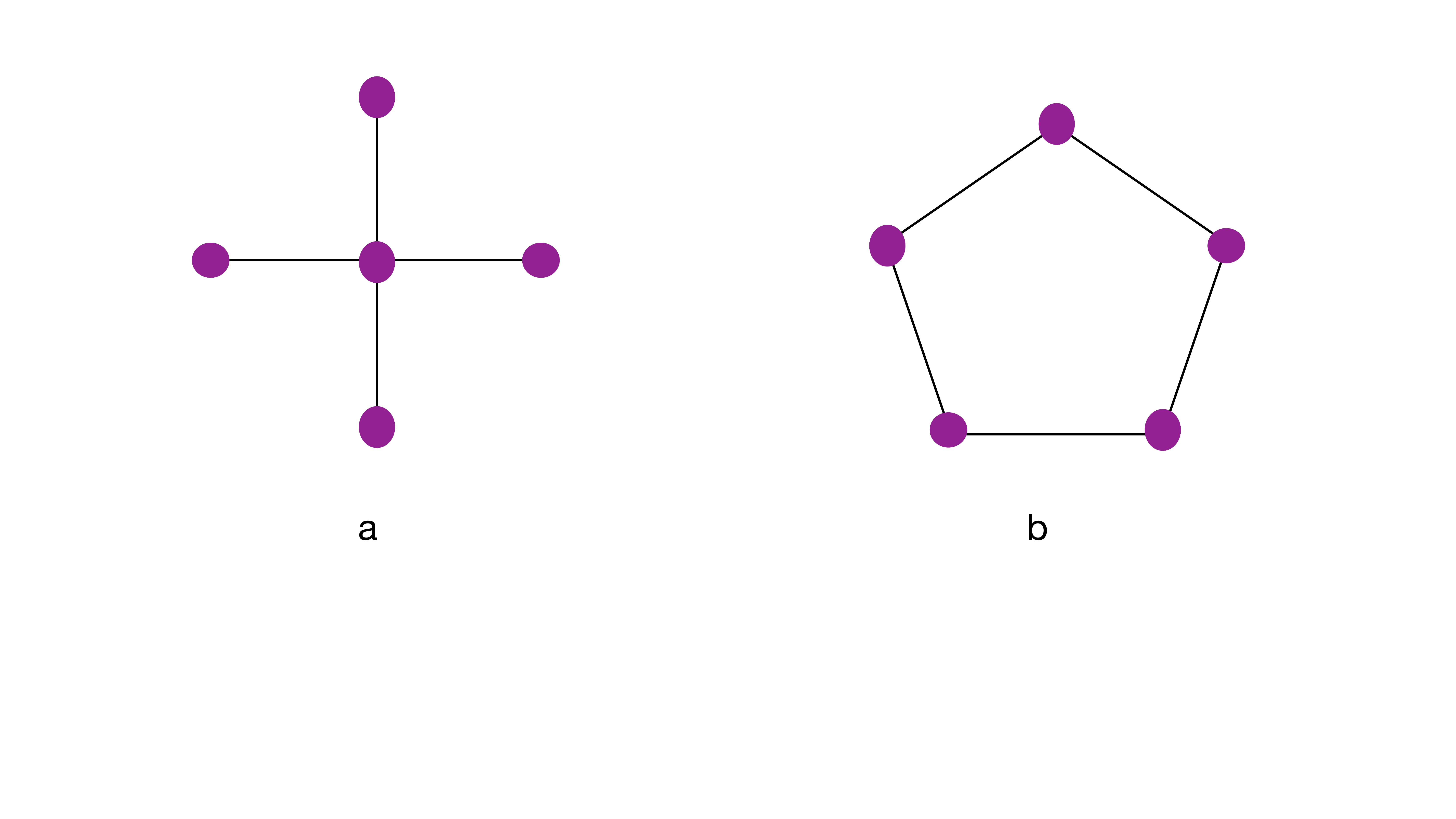}\vspace{-2.5cm}
	\caption{A PME state of five qudits can correspond to many different graphs with different connectivity structures. Two different examples are shown here. In this work we consider the simplest ring geometry for all number of qudits.  }\label{fig5}
\end{figure}

\begin{figure}[!ht] 
	\centering
	\includegraphics[scale=0.5]{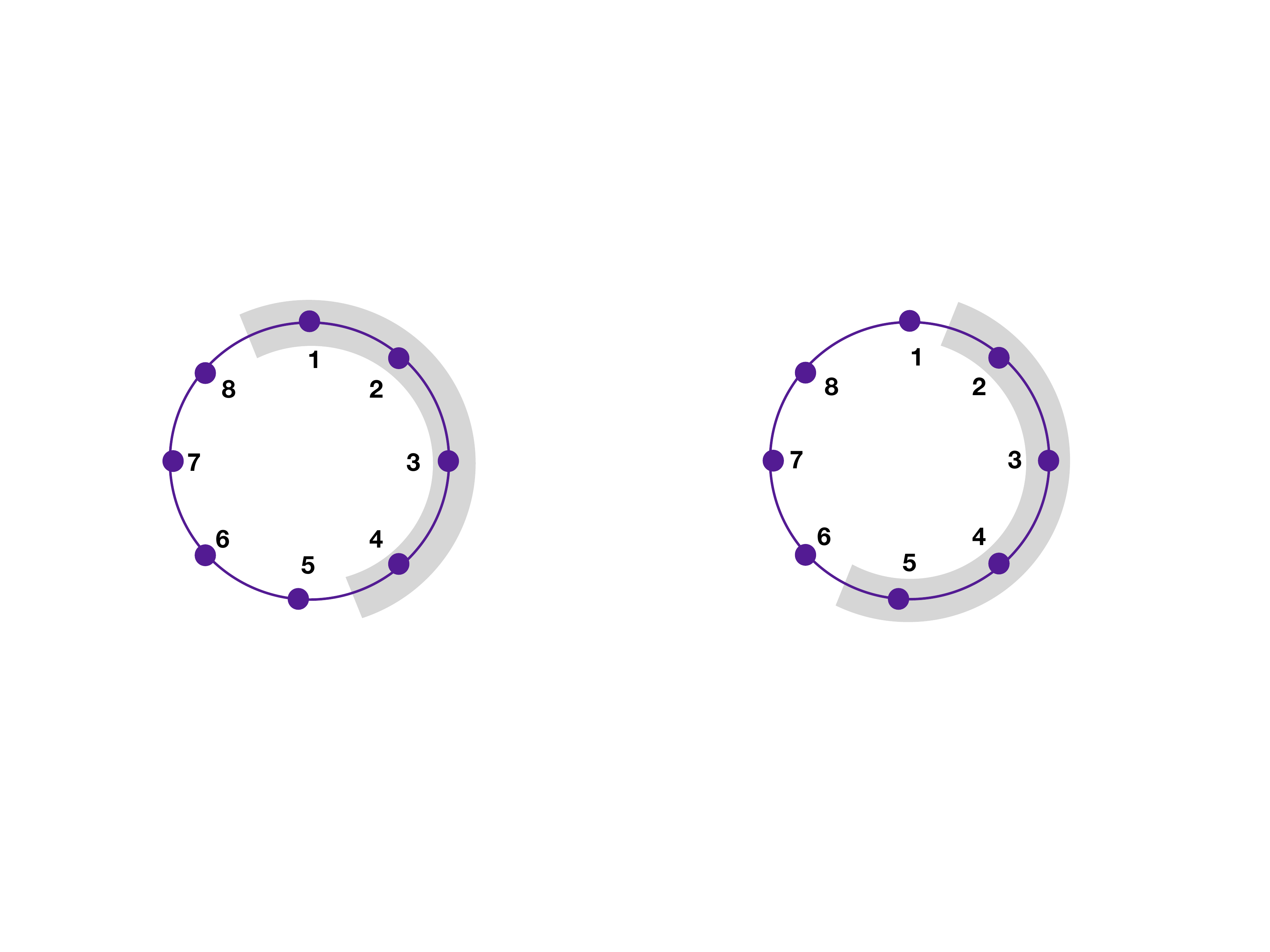}\vspace{-4.0cm}
	\caption{An eight-particle state is a PME if for any partition consisting of four adjacent particles (the shaded area) the density matrix is proportional to identity. }\label{PMEPartition}
\end{figure}

\noindent  {\bf Remark on the  geometry:} The notion of connected subsets, naturally brings up the concept of geometry and connectivity of a graph corresponding to the state. This may not have been so explicit in this section when the case of four particles was considered but is certainly relevant for higher number of particles. For example consider the next simple case of 5-particle states, where as an example we show two  different  underlying geometries  as in figure (\ref{fig5}).  Certainly the requirement of connected subsets of $n\leq 2$ particles being in a completely mixed state depends on which graph we have in mind. In this work we always consider the simplest one dimensional geometry with open or periodic boundary conditions.    The question of existence and construction of PME states on arbitrary graphs is an interesting but highly non-trivial problem which can be investigated in the future. \\

\section{Examples of PME states in any dimension }
\label{pmeN}

\noindent  A  state $$\ket{\Psi} =\frac{1}{\sqrt{d^{|A|}}} \sum_{A,A^c}\Psi_{A^c,A}\ket{A^c}\ket{A}$$ of $|A+A^c|$ qudits 
is a PME state if for any division of the qudits into connected subsets of equal size $|A|=|A^c|$, the operator 
$$\hat{\Psi} = \sum_{A,A^c}[\Psi]_{A^c}^{A}\ket{A^c}\bra{A},$$
is unitary, in which 
$$[\Psi]_{A^c}^{A}=\Psi_{A^c,A}$$
is the matrix representation of the state coefficients.  More specifically 
a state of $2n$ qudits like 
\be
|\Psi\ra=\frac{1}{\sqrt{d^{n}}}\Psi_{i_1\cdots i_{2n}}|i_1,i_2,\cdots i_{2n}\ra ,
\ee
is a PME state if the following matrices constructed from these coefficients are all unitary
\be
\label{PMEcond}
[\Psi]_{i_1...i_n}^{i_{n+1}...i_{2n}} = [\Psi]_{i_2...i_{n+1}}^{i_{n+2}...i_{2n}i_1}= \cdots = [\Psi]_{i_n...i_{2n-1}}^{i_{2n}...i_{n-1}} .
\ee

\noindent It is then obvious that constructing PME states for arbitrary number of qudits is a highly non-trivial task, let alone their classification. 
In this section we give some simple examples of these states  for an even number of parties $2n$. These simple examples  can then be turned into rather large multi-parameter families by appropriate action of quantum gates,  as we will see.\\

\noindent Let $|\phi^+\ra$ be a maximally entangled state of the form 
\be
|\phi^+\ra=\frac{1}{\sqrt{d}}\sum_{i=0}^{d-1} |i,i\ra.
\ee
Then the state 
\be\label{PhiBell}
|\Phi^+\ra:=\prod_{k=1}^n|\phi^+\ra_{k,k+n}
\ee
shared between $2n$ parties is a PME state. \\

\noindent {\bf Remark:} All our results in this section are valid for arbitrary $n$, however in order not to clutter the text with multiple subscripts and to stress only on the main ideas we explain only the case $n=4$, i.e. states with $8$ qudits. It is readily seen by looking at the formulas and the graphs, that the arguments are also valid for general $n$. \\

\begin{figure}[!ht] 
	\centering
	\includegraphics[scale=0.5]{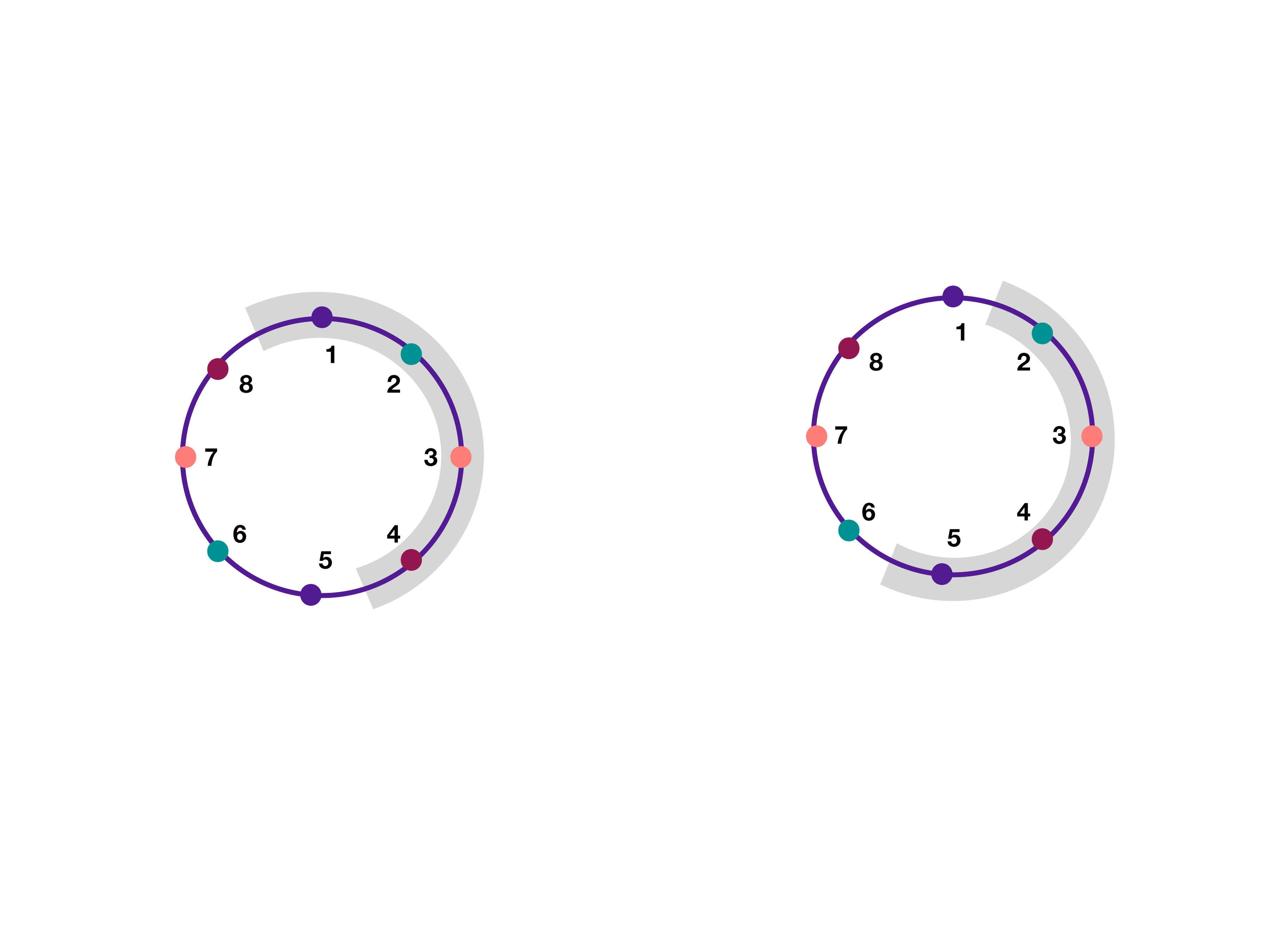}\vspace{-4.0cm}
	\caption{This figure shows clearly why states of the form (\ref{PhiBell}) are PME. Bell pairs are on the end points of diagonals. Any subset of adjacent particles, shaded area, contain one-half of a Bell pair which is in a completely mixed states.}\label{PMEBell}
\end{figure}

\noindent  The proof is by expansion as seen by a simple example, say for $n=4$. The state is then given by 
\be\label{Belldimer}
|\Phi^+\ra=\frac{1}{d^2}\sum_{i,j,k,l}|i,j,k,l,i,j,k,l\ra.
\ee
It is then clear that any subsequent indexes for four particles pertain to a complete set of orthogonal basis with the remaining indices pertaining to another similar set. This is also best seen in figure (\ref{PMEBell}). 
\noindent One can easily check that any two consecutive qudits in this state are orthogonal, proving that the state is a PME. 
We call these states dimerized PMEs, since they are a juxtaposition of dimer states of $|\phi^+\ra$ on sites of a chain, although each dimer is extended over sites which are far away. \\

\noindent We can now construct more complicated PMEs by acting on the state $|\Psi\ra$,  by controlled operators. Let 
\be
|{\Psi}\ra=\Lambda_{n+1,n+2}(U_1)\Lambda_{n+2,n+3}(U_2)\cdots\Lambda_{2n-1,2n}(U_{n-1})|\Phi^+\ra,
\ee
where $\Lambda_{i,j}(U)$ is the controlled operator acting on sites $i$ and $j$, where site $i$ is the control site and site $j$ is the target site:
\be
\Lambda(U)|i\ra\otimes |j\ra=|i\ra\otimes U^{i}|j\ra=:|i\ra\otimes |U(i,j)\ra,\h i,j=0,\cdots d-1,
\ee
where in the last equality we have introduced the shorthand notation 
\be
|U(i, j)\ra=(U)^i|j\ra.
\ee
This is a class of multi-parameter PME states. For the sake of clarity, we write their explicit form for the case when $n$ is a  
 small number, say $n=4$. The argument for this simple case can easily be seen to work for arbitrary values of $n$. For $n=4$, we have
\ba\label{4U}
|{\Psi}\ra&=&\Lambda_{5,6}(U)\Lambda_{6,7}(V)\Lambda_{7,8}(W)|\Phi^+\ra\cr &=&{\frac{1}{d^2}\sum_{i,j,k,l}|i,j,k,l,i,U(i,j),V(j,k),W(k,l)\ra}.
\label{twoEntryPME}
\ea
It is obvious that
\be\label{Uortho}
\la U(i,j)|U(i,j')\ra=\delta_{j,j'},
\ee
with similar results when $U$ is replaced by any other unitary operator. 
We should now prove that any four consecutive index set in the expansion (\ref{4U}), corresponds to a full set of orthogonal vectors. A direct examination of (\ref{4U}) using the result (\ref{Uortho}) shows that this is indeed the case. As an example consider the indexes $3,4,5,6$. Starting from the leftmost states and using (\ref{Uortho}) repeatedly, we find
\be
\la l,i,U(i,j),V(j,k)|l',i',U(i',j'),V(j',k')\ra=\delta_{l,l'}\delta_{i,i'}\delta_{j,j'}\delta_{k,k'},
\ee
which clearly shows that the density matrix of the rest of the particles is proportional to identity.\\

\begin{figure}[!ht] 
	\centering
	\includegraphics[scale=0.3]{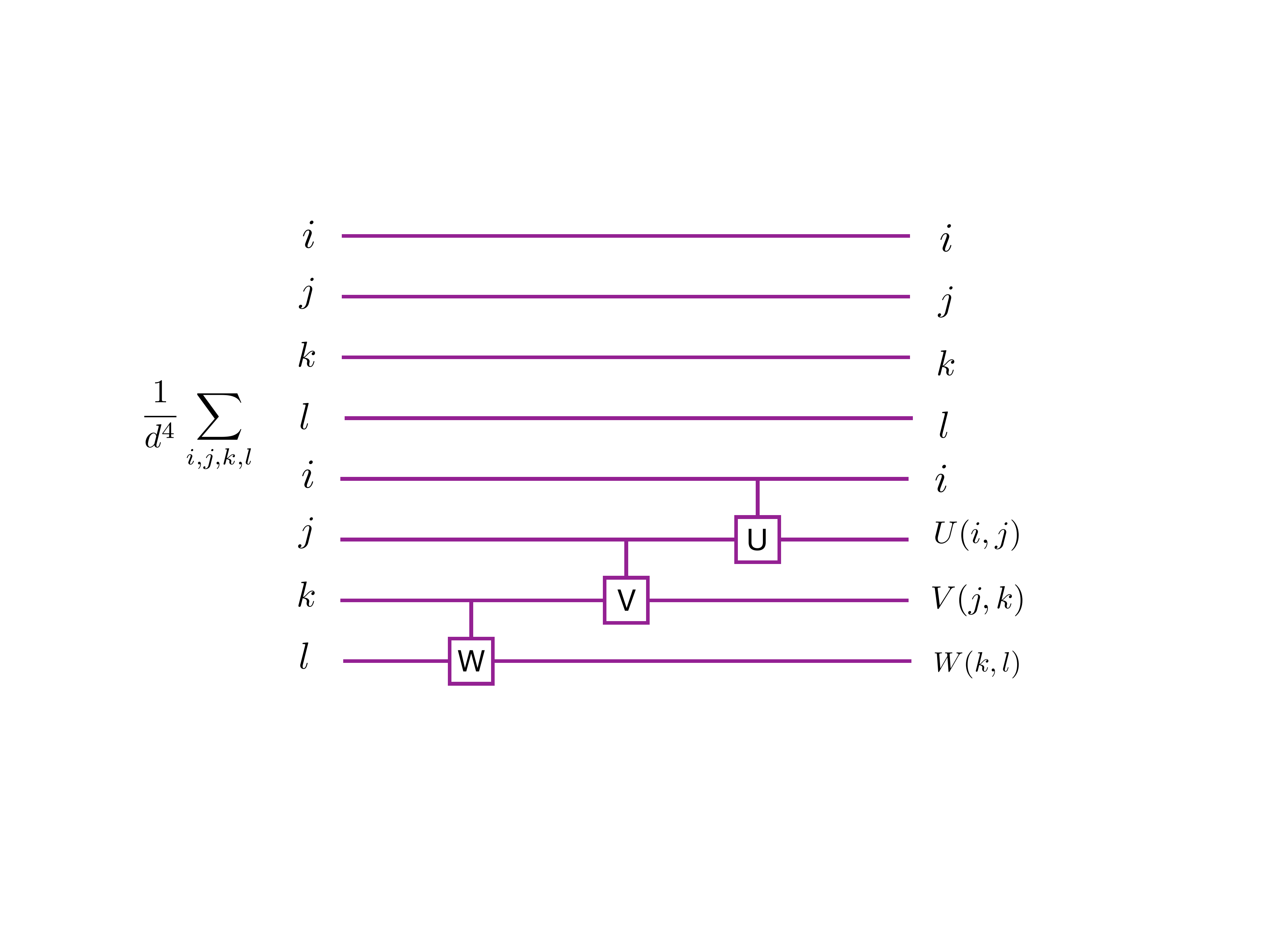}\vspace{-1.5cm}
	\caption{The quantum circuit for producing the state $|\Psi\ra$ for 8 particles. The input is the state $|\Phi^+\ra$ in Eq. (\ref{Belldimer}).}\label{PMELambda}
\end{figure}
\noindent  Figure (\ref{PMELambda}) shows the circuit which produces these states when the input state is the basic  PME state $|\Phi^+\ra$ of equation (\ref{Belldimer}). \\

\begin{figure}[!ht] 
	\centering
	\includegraphics[scale=0.3]{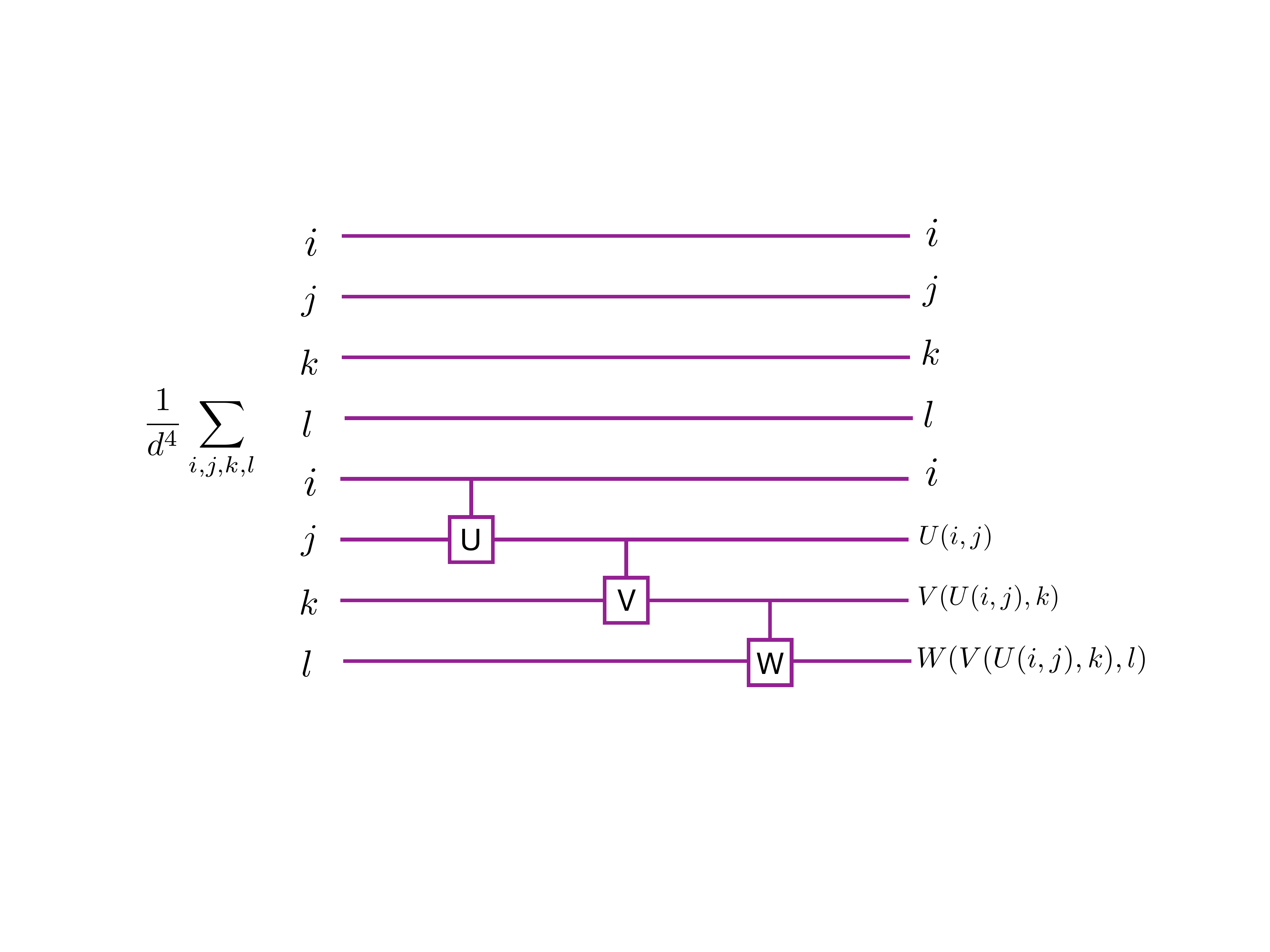}\vspace{-1.5cm}
	\caption{The quantum circuit for producing the state $|\Xi\ra$ for 8 particles. The input is the state $|\Phi^+\ra$ in Eq. (\ref{Belldimer}).}\label{PMELambda2}
\end{figure}

\noindent One can produce another class of PME states 
and
\be
|{\Xi}\ra=\Lambda_{2n-1,2n}(U_{n-1})\cdots \Lambda_{n+2,n+3}(U_2) \Lambda_{n+1,n+2}(U_1)|\Phi^+\ra,
\ee
by reordering the controlled gates of the circuit in figure (\ref{PMELambda}), as shown in figure (\ref{PMELambda2}). 
The explicit form of the resulting state (for $n=4$) is now written as:
\ba\label{}
|\Xi\ra&=&
\Lambda_{7,8}(W)\Lambda_{6,7}(V)\Lambda_{5,6}(U)|\Phi^+\ra\cr &=&
{\frac{1}{d^2}\sum_{i,j,k,l}|i,j,k,l,i,U(i,j),V(U(i,j),k),W(V(U((i,j),k)),l)\ra}.
\ea

\noindent  {The proof that these states are PMEs is exactly like the one presented above for the states (\ref{4U}). }

 \noindent Note that as long as the operators $U,\ V$ and $W$ do not commute, the two circuits in figures (\ref{PMELambda}) and (\ref{PMELambda2}) cannot be converted to each other and they produce inequivalent PME states. 
 
\section{Applications of  PME states}\label{pmeApp}
\noindent It has been shown that AME states, due to their nice properties, can find applications in many areas, notably in parallel teleportation and quantum secret sharing \cite{helwig2012absolute} and in error correcting codes \cite{raissi1} and finally in making microscopic models of holographic duality \cite{Almheiri_2015,harlow2018tasi,pastawski2015holographic}. Naturally PME states, which share many of the properties of AME states may also find such applications to a lesser or even greater degree. We will now explore these aspects. 

\subsection{Parallel Teleportation}

\noindent Given a PME state, if we split the qudits into two planar groups $A$ and $B$ (i.e. where each group is a connected partition)  with $|A|=|B|$, we have seen that the PME can be written as 
\begin{equation}
	\ket{\psi_{A,B}} = \frac{1}{d^{m/2}}\sum_{K} \ket{K}_AU(\ket{K}_B),
\end{equation}
where $U$ is a unitary and $\{|K\ra\}$ is a basis for the $A$ or the $B$ part.  
\noindent So, if the $B$ part performs a unitary transformation $U^{\dagger}$ on their qudits, the state turns into 
\begin{equation}
\ket{\phi_{A,B}} = \frac{1}{d^{m/2}}\sum_{K} \ket{K}_A\ket{K}_B,
\end{equation}
which is a maximally entangled state. It is now possible for A and B to teleport any state of dimenions $d^{|A|}$. 
Note that this requires that the $B$ part be all in one place or lab, such that a unitary  $\in SU(|B|)$ can be applied to that part of the state.  The main difference between an AME and a PME in this case is that for an AME, teleportation is possible between any two partitions, while here it is possible only between connected partitions. 

\subsection{Quantum State Sharing with PME states}
\noindent The PME states  can be used for quantum state sharing (QSS) between $2n-1$ players so that any adjacent players of size greater than $n$ can recover the state. The difference with the AME state is that the players should be adjacent to each other, i.e. they should form a connected subset of the set of players. The idea can be understood by a simple example. Consider the case $n=2$ where the  PME state is
\be
|\Psi\ra=\sum_{i,j,k,l}\psi_{i,j,k,l}|i,j,k,l\ra_{1,2,3,4}.
\ee
In view of Eq. (\ref{basicPsi}), the  PME state of four particles can be expressed in different ways, depending on which bi-partition we are interested it as
\be
|\Psi\ra=\frac{1}{d}\sum_{i,j}|i,j\ra_{1,2}\otimes U|i,j\ra_{3,4},
\ee
or equivalently 
\be
|\Psi\ra=\frac{1}{d}\sum_{i,j}|i,j\ra_{1,4}\otimes V|i,j\ra_{2,3},
\ee
where $U$ and $V$ are two unitary operators.
Let the first player be the distributer whom we call Alice. She can then encode a basis state as
\be
|i\ra\lo |S(i)\ra:=\frac{1}{\sqrt{d}}\sum_{j}|j\ra_{2}\otimes U|i,j\ra_{3,4}=\frac{1}{\sqrt{d}}\sum_{j} V|i,j\ra_{2,3}\otimes |j\ra_{4}.
\ee
An arbitrary state is obviously encoded by linear extension 
\be
|\a\ra=\sum_i \a_i|i\ra\lo |S(\a)\ra=\sum_i \a_i|S(i)\ra. 
\ee
If the players $(3,4)$ want to retrieve the state, they  act on the state $|S(i)\ra$ by the operator $U^{-1}$ to decode it to  
\be
({U^{-1}})_{3,4}|S(i)\ra\lo \frac{1}{\sqrt{d}}\sum_{i,j} |j,j\ra_{2,4}|i\ra_3=|\phi^+\ra_{2,4}|i\ra_3,
\ee
and if the players $(2,3)$ want to retrieve the state, they  act on the state $|S(i)\ra$ by the operator $V^{-1}$ to decode it to 
\be
({V^{-1}})_{2,3}|S(i)\ra\lo \frac{1}{\sqrt{d}}\sum_{i,j} |j,j\ra_{3,4}|i\ra_2=|\phi^+\ra_{3,4}|i\ra_2.
\ee
In both cases an arbitrary state is retrieved by one member of the legitimate party. The same reasoning is true for other legitimate sets, like (4,1). \\

\begin{figure}[!ht] 
	\centering
	\includegraphics[scale=0.4]{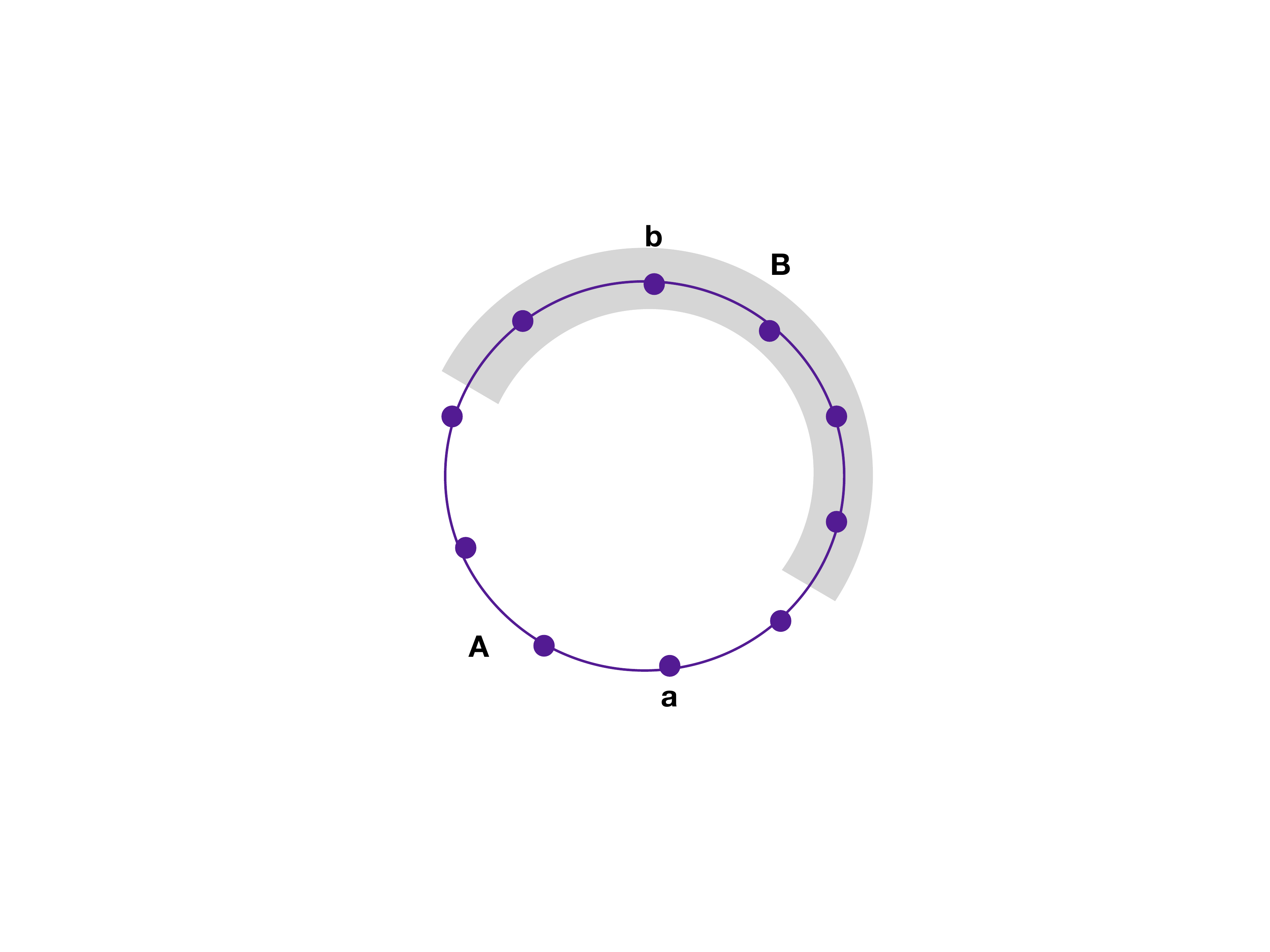}\vspace{-2.5cm}
	\caption{The distributer of the secret state is $a$, who encodes the state $|\a\ra$ and sends it to all of the other players. Any connected subset (b,B) of size $n$ or larger can collaborate so that $b$ can retrieve the state. }\label{QSSRing}
\end{figure}

\noindent This argument can easily  be extended to larger number of players. Let us denote the distributor by the index $a$, call her Alice, the player who is going to retrieve the state by $b$, call him Bob. The collaborators of Bob form a connected subset of size $n-1$. Call this subset $B$. (The shaded subset in figure (\ref{QSSRing}) includes both Bob and $B$.)

\noindent The rest of the players together  form a connected subset of size $n-1$ which we call $A$.  \\

\noindent Then the state can be written as 
\be
|\Psi\ra=\frac{1}{\sqrt{d}^n}\sum_{k_0,{\bf k}}|k_0,{\bf k}\ra_{a,A}\otimes |\Phi(k_0,{\bf k})\ra_{b,B} =\frac{1}{\sqrt{d}^n}\sum_{k_0,{\bf k}}|k_0,{\bf k}\ra_{a,A}\otimes U_{A}|k_0,{\bf k}\ra_{b,B},
\ee
where we have used the PME property that for any connected subset of size $n$, $\la \Phi(k'_0,{\bf k'})|\Phi(k_0,{\bf k})\ra =\delta_{k_0,k'_0}\delta_{{\bf k},{\bf k'}}$. Now Alice encodes the qudit 
	$|k_0\ra_a$ as follows
\be
|k_0\ra_a\lo |S(k_0)\ra_{b,A,B}=\frac{1}{\sqrt{d}^{n-1}}\sum_{{\bf k}}|{\bf k}\ra_{A}\otimes U|k_0,{\bf k}\ra_{b,B}.
\ee
We now note that the $A$ part is in a completely mixed state, since
\be
\rho_A=Tr_{b,B}( |S(k_0)\ra\la S(k_0)|)=\sum_{{\bf k}}|{\bf k}\ra\la {\bf k}|=\frac{1}{d^{n-1}}I_A,
\ee
which means that neither the $A$ nor any smaller subset of it can recover the state. Furthermore, we find that the state can be retrieved by members in $(b,B)$ by applying ${U}^{-1}$ 
\be
{U}^{-1}|S(k_0)\ra_{b,A,B}=\frac{1}{\sqrt{d}^{n-1}}\sum_{{\bf k}}|{\bf k}\ra_{A}\otimes |k_0,{\bf k}\ra_{b,B}=|k_0\ra\otimes \frac{1}{\sqrt{d}^{n-1}}\sum_{{\bf k}}|{\bf k}\ra_{A}\otimes |{\bf k}\ra_{B}=|k_0\ra_b\otimes |\Psi\ra.
\ee
{\bf Remark:} Note that this division of players is completely arbitrary as long as the size of the $A$ and the  $B$ part are equal to $n$. The form of the operator $U$ depends on the bi-partition, as in Eqs. (\ref{UUU} and \ref{VVV}), but we are sure that such a unitary exists due to the PME property of the state. 
\\

\noindent In summary, like AME states, PME states can be used for quantum state sharing and any subset of players of size larger than half of the players  can  recover the encoded state, as shown above, by acting appropriately on their share of the state. The only condition is that this subset should be a connected subset. 

\section{ Conclusions}
\noindent Planar Maximally Entangled (PME) states are those n-partite states for which each set of {\it{adjacent}} particles whose size is less than or equal to $\floor*{\frac{n}{2}} $ is in a totally mixed states. These are a generalization of Absolutely Maximally Mixed (AME) states which have been extensively studied and have many applications in quantum information theory. The AME states are a subclass of PME states, that is, any AME state is a PME state, but the converse is not true. 
We have investigated these states, classified all such states for four qubits, where we have shown that there are two families of such states, a two-parameter family and a four parameter family, the intersection of which is the dimerized Bell state $|\Psi\ra=|\phi^+\ra_{13}|\phi^+\ra_{24}$. This is in contrast to AME states where there is no such state \cite{higuchi}. We have then constructed a large family of  such states in any dimension and for any even number of particles. The PME states share many of the properties of AME states which makes them suitable for application in parallel teleportation, in quantum secret sharing and in quantum error correction, with appropriate modification. These modifications are such that "any subset" in AME applications should generally be replaced by "connected subsets" or subsets of adjacent particles provided that their size exceeds the threshold of 
$\floor*{\frac{n}{2}} $. \\

\noindent  Although we have constructed large families of PME states, this construction may not be exhaustive and it would be interesting to classify all possible PME states for any number of particles and any dimension.
Another interesting line of research is to see if or how PME states can replace the AME states in constructing microscopic models of Holographic duality and AdS/CFT correspondence. It is known that some kinds of AME states, like the 5 qubit code, have been used in constructing such models known HaPPY code \cite{pastawski2015holographic}. The main property which has been effective in constructing such models is the correspondence between AME states and isometries and the graphical representation of tensors. In the AME case the isometric property holds true no matter how we divide the legs of a tensor into input and output legs. In the PME case this division can be made only for adjacent legs. Whether these kind of tensors can be used to make tessellations of two-dimensional space is an interesting line of work.

\section{Acknowledgements} MD would like to thank Ipak Fadakar, Ali Naseh, Behrad Taghavi and Hamed Zolfi for fruitful comments and discussions. The authors thank V. Jannessary, A. Mani, M. Moradi, K. Sadri, F. Shahbeigi,  and M. Taheri  for their constructive comments. This project was partially supported by the grant G950222 from Sharif University of Technology.


\end{document}